\documentclass[twocolumn,showpacs,preprintnumbers,floatfix,prl]{revtex4}

\usepackage{graphicx}

\begin{document}

\title{Effect of friction in a toy model of granular compaction}
\author{F. Ludewig, S. Dorbolo and N. Vandewalle}
\affiliation{GRASP, Institut de Physique B5, Universit\'e de Li\`ege, \\ B-4000 Li\`ege, Belgium.}

\begin{abstract}
\end{abstract}

\begin{abstract}
We proposed a toy model of granular compaction which includes some resistance due to granular arches. In this model, the solid/solid friction of contacting grains is a key parameter and a slipping threshold $w_c$ is defined.  Realistic compaction behaviors have been obtained.  Two regimes separated by a critical point $w_c^*$ of the slipping threshold have been emphasized : (i) a slow compaction with lots of paralyzed regions, and (ii) an inverse logarithmic dynamics with a power law scaling of grain mobility. Below  the critical point $w_c^*$, the physical properties of this frozen system become independent of $w_c$. Above the critical point $w_c^*$, i.e. for low friction values, the packing properties behave as described by the classical Janssen theory for silos. 
\end{abstract}

\maketitle        

\section{Introduction}

In our industrial world, most of the products are processed, transported, and stocked in their granular state. The density of those granular systems appears therefore to be a crucial parameter for evident economic reasons. In this spirit, the physics of compaction is thus relevant for a broad range of applications \cite{herrmann,behringer}.

Some experimental studies report the evolution of granular packing submitted to successive taps. Experiments on vibrated granular materials exhibit low compaction \cite{knight} and phase segregation \cite{kudrolli}. In most cases, the density of the material slowly changes as a function of the tap number $t$ and an inverse logarithmic dynamics has been observed. For a few experiments, a fast exponential saturation of the density is obtained \cite{epje}. Also, some granular materials may present small density variations (about $1\%$) from loose packing to dense configurations while, for others, high variations may be observed (up to $20\%$ \cite{geo}). Therefore, granular materials exhibit a large diversity of compaction behaviors : according to the taps (reduced acceleration $\Gamma$ and numbers) and according to the ability to pack (nature and shape of the grains). It is of interest to find out the physical parameters that are relevant for the occurrence of those different compaction dynamics.

In order to reproduce a slow compaction dynamics, the so-called Tetris model has been introduced by Caglioti and coworkers \cite{caglioti}. This toy model considers rectangular blocks placed on a square lattice tilted by $45^{\circ}$. The blocks are submitted to gravity and cannot overlap. At each simulation step, the entire packing is perturbated by a virtual tap in which  a number $n$ of motoins per grain are realized. During each dynamic step, a grain is moved upward with a probability $p_{up}$ (0 $<$ $p_{up}$ $<$ $\frac{1}{2}$) and downward with a probability ${p}_{down} = 1 - p_{up}$. In the same time, the grain can rotate with a probability  ${p}_{rot}$ (= $\frac{1}{2}$). After the dynamical step, the packing relaxes until all grains reach an equilibrium position with respect to gravitation. The main parameter of the Tetris model is the probability ratio $p_{up}/(1-p_{up})$. As a function of this parameter, the acceleration of the taps are tuned and some inverse logarithmic behaviors emerge, for small values. Recently, the Tetris model has shown some relevance for the study of phase segregation occurring in granular binary mixtures \cite{nicodemi}. 

Our main motivation is to propose a simple model which reproduces the behaviors of compaction dynamics and which includes the effect of friction.

\section{Compaction Model}

\begin{figure}[h] 
\begin{center} 
\includegraphics[width=6cm]{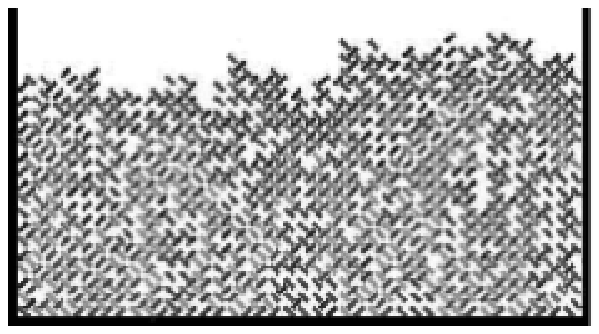}
\includegraphics[width=6cm]{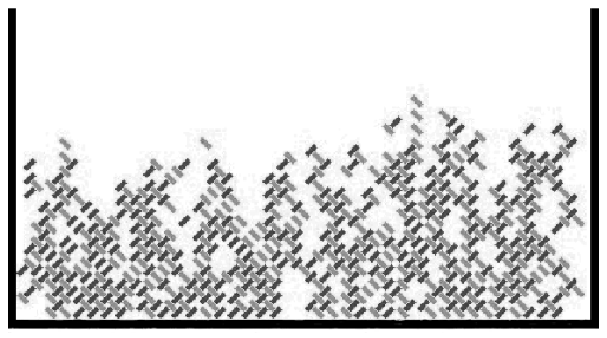}
\caption{(top) Initial configuration of a packing obtained by the `rain' method (see test) for $w_{c} = 12$.  The gray scale levels emphasize the orientation of the grain.  (bottom) Only grains that weight is larger than $w_{c} = 12$ are represented.} 
\end{center} 
\end{figure}

Our toy model is the following. The grains are represented by rectangular blocks, as in the original Tetris model.  Two orientations are possible on the tilted lattice.  The main rule of this model is a geometrical constraint : two adjacent blocks cannot have the same orientation along the principal axis of the blocks.  In so doing, the close packing is realized when lines of blocks of one orientation alternate with lines of blocks of the other one, i.e. when the system reaches a quincunx configuration.  Along this paper, we will refer to this packing as the well-ordered packing or ground state. The potential energy is the lowest. In this state, there is no hole in the structure and a flat top surface.

The initial packing is built using a `rain' method : the grains are dropped from a random position on the top and stopped when they reach the heap.  A loose-packed heap is obtained.  A typical configuration is shown in Figure 1 (top).

In order to take into account the friction, the apparent block weights $w_i$ are introduced for every block $i$ with a slipping threshold.  The weigth of a single block $w_b$ is assumed to be equal to 1 in arbitrary units.  Since this calculation is completely deterministic and depends on the block orientations and positions, the contact forces are calculated at each step for every blocks from top to bottom. The additional rule for including friction is based on a More-Coulomb criterium \cite{livre} : when a grain has an apparent weight $w_i$ larger than a given value $w_c$, it cannot move.  Figure 1 (right) shows the blocks for which the apparent weights are greater than $w_c = 12$.  This slipping threshold $w_c$ has a physical relevance.  Indeed, the friction force $F_i$ between two grains is given by a certain fraction $\mu_s$ of the normal force at the contact. The larger $w_i$, strongest is the friction force.  If a grain has a weight larger than $w_c$, the friction is such that it avoids any movement of the grain due to the shock produced by the tap.  Thus, the parameter $w_c$ allows us to tune the amplitude of taps $or$ the inverse of the static friction $\mu_s$.  Note that the asymptotic value $w_c \rightarrow \infty$ reduces the model to the Tetris one.  On the other hand, a slipping threshold value $w_c=1$ leads to a frozen situation, namely when friction forces are larger than vibration ones (or that the grains are glued together). 

In this work, borders have been considered in order to evidence some redirection of the forces towards the walls.  The friction between a grain and the border is the same as between two grains.

The rules of a tap simulation are nearly the same as for the original Tetris model.  It can be described in two steps : (i) excitation and (ii) relaxation.  

(i) A number $N$ of grains are randomly selected. In our simulation, this number $N$ is fixed to the total number of grains lying on the lattice.  When the geometry allows it, the grain goes up with a probability $p_{up}$ or goes down with a probability $p_{down} = 1-p_{up}$ . After grain's translation, if the selected grain has at least 3 free neighboring sites, it can rotate with a probability 1/2.  

(ii) The entire system is then relaxed. The grains go down until no grain motion is allowed downward. The only one driving force is the gravity.
\begin{figure}[h]
\begin{center} 
\includegraphics[width=6cm]{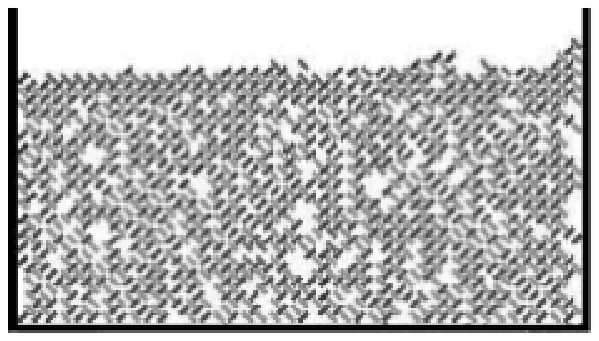}
\includegraphics[width=6cm]{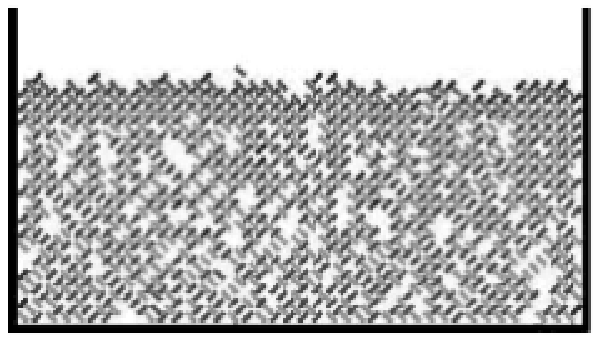}
\includegraphics[width=6cm]{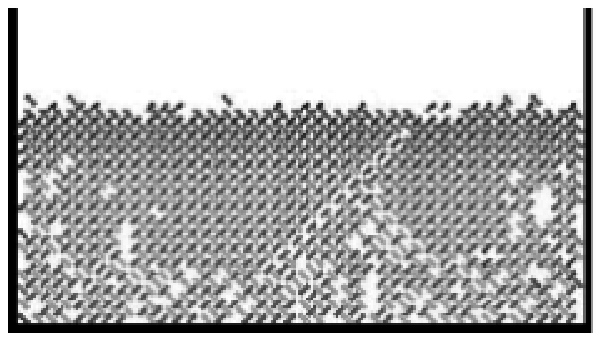}
\includegraphics[width=6cm]{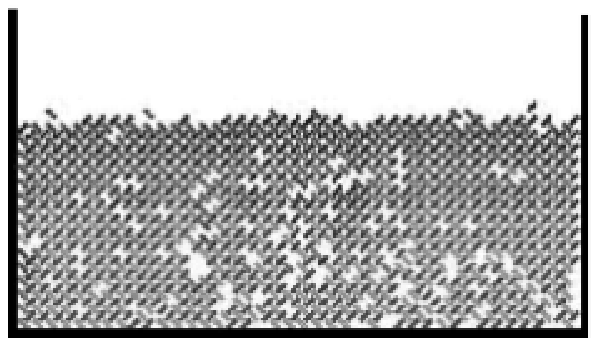}
\includegraphics[width=6cm]{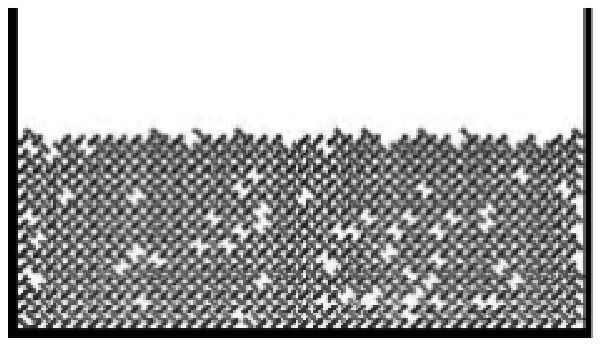}
\caption{Simulation of toy model.  Six obtained configurations of the heap for different thresholds $w_c$ after $10^4$ MCS taps.  From top to bottom : $w_c=$2, 4, 12, 16 and 64.} 
\end{center} 
\end{figure}

\section{Results}

Simulations have been performed for various square lattice sizes from $L$ = 40 to $L$ = 120,. On such lattices, no jamming is observed. The average total number of particles placed on the lattice with the rain method is equal to two third of the total free sites. The value of $p_{up}/(1-p_{up})$ has been fixed to 1. Changing this parameter will not modify qualitatively the results discussed below. 

\subsection{Evolution of the density $\rho$}

The density $\rho$ is evaluated by dividing the total number of the sites by the number of particules in the lowest 25 $\%$ of the lattice. Figure 2 presents the compacted heap after $10^4$ Monte Carlo Step (MCS) taps for a 40 $\times$ 40 lattice. The slipping threshold $w_c$ has been set to 2, 4, 12, 16, 32, 64 from top to bottom.  The situation $w_c=2$ is the most constricting one.  Indeed, when a block has an upper neighbor, its apparent weight is equal to 2 : this block is immediately paralized. For a greater value of the slipping threshold $w_c$, more grains are free to move. The larger value of $w_c$ investigated herein is $w_c = 6400$, it should correspond to a model without friction (like the Tetris one).  

Two kinds of structures can be seen across the packing : force chains and cavities.  The chains of force are defined by the blocks which have a high effective weight.  They form lines that propagate through the whole packing.  These chains are responsible for the redistribution of the weights in the packing. Moreover, they can lean on the borders, thus a part of the total weight is supported by the borders. Subsequently, some blocks are allowed to support less weight even when they are located near the bottom of the packing.  Below the chains of force, caverns may then be found.  They are holes in the packing which decrease the global density of the packing because arches resist to the taps.

Caverns are particularly visible in Figure 2. As $w_c$ increases, the size of the caverns decreases.  Indeed, for large $w_c$ values, the caverns are reduced to point defects in the `crystalline' quincunx structure.  Those defects are very stable, especially when they are located in the bulk of the heap.  Among those stable defects, a hole can be surrounded by four blocks which are oriented towards it.  Such singularities come from the geometrical laws.  

The packing is of course closely linked to the number of these caverns inside the packing.  It is noticeable that the number of caverns increases with the considered depth of a packing.  In other words, the well-ordered phase {\it grows} from the top of the packing since the grains are less stressed there.  This phenomenon is particularly well seen in Figure 2 for the particular case $w_c=12$.  The notion of cavern is also related to the formation of arches.

The comparison of the heaps after a compaction process shows that the higher $w_c$ is, the more packed the system becomes. In so doing, the compaction $\rho$ has been plotted with respect to the number of taps in Figure 3.  The graph is represented in a semi-log scale.  Several values of the slipping threshold $w_c$ have been tested and are represented by different symbols (see the legend).  The typical evolution of the density is slow during the 10 first taps before it drastically increases.  A saturation occurs towards a maximum value  $\rho_{\infty}$.  The solid curves correspond to fits by a three-parameters law \cite{knight, caglioti}
\begin{equation}
\label{tanh}
\rho(t) = \rho_{\infty}  - \frac{\Delta\rho}{ 1 + B\ln(1 + t/\tau)}
\end{equation} 
where $ \rho_{\infty}$ is the asymptotic density when $t\rightarrow\infty$, $\Delta\rho$ is the maximum variation of the density, $B$ is a free fitting parameter and $t$ is the number of taps. The parameter $\tau$ is a characteristic number of taps. The inverse logarithmic behavior Eq.(\ref{tanh}) is relevant for a slow dynamics of compaction. The complete evolution of $\rho$ can be fitted by this unique law Eq.(\ref{tanh}) for all $w_c$ values.

\begin{figure}[htbp]
\begin{center} 
  \includegraphics[width=8cm]{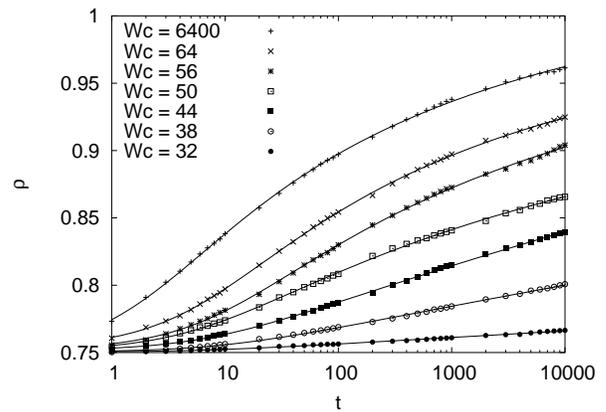}
\caption{Evolution of the density $\rho$ with respect to the tap number $t$ in a semi-log plot.  The different labels correspond to different slipping threshold $w_c$ values which are indicated in the legend. From bottom to top : $w_c =$ 32, 38, 44, 50, 56, 64 and 6400. Simulations have been performed on 100 $\times$ 100 lattices. The solid curves are fits using the inverse logarithmic law Eq.(\ref{tanh}).} 
\end{center} 
\end{figure}

\begin{figure}[htpb] 
\begin{center} 
\includegraphics[width=8cm]{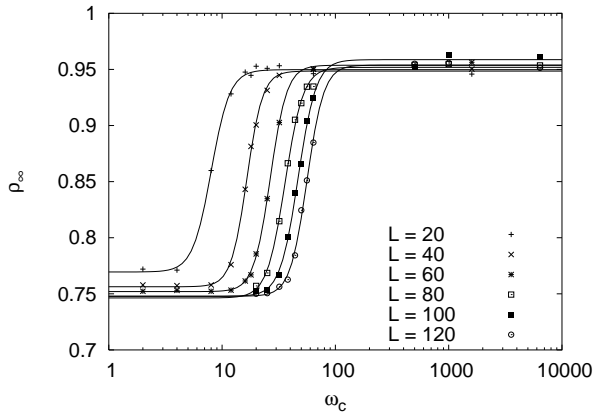}
\caption{ Semi-log plot of the asymptotic density $\rho_{\infty}$ as a function of the slipping threshold $w_c$ for different lattice sizes : $L$ = 20, 40, 60, 80, 100 and 120. The solid curves are fitted hyperbolic tangent for determining the critical value $w_c^*$ of the slipping threshold $w_c$.} 
\end{center} 
\end{figure}

 The form of Eq.(\ref{tanh}) allows us to determine the saturation density $\rho_{\infty}$ by fitting.  The values of $\rho_{\infty}$  are reported in Figure 4, with respect to the slipping threshold $w_c$.  The different curves are obtained for different lattice sizes $L$ : from  $L =$ 20 up to $L =$ 120 lattices. The asymptotic behavior depends on the slipping threshold $w_c$ : the higher the threshold is, higher the saturation is. Whatever the size of the lattice, the maximum density $\rho_{\infty}$ is obtained for high values of the slipping threshold and is equal to approximatively 0.96.  A higher density around 1 would have been expected but the presence of the remaining defects are responsible for this difference. An inflexion point is found whatever the considered size of the heap.  Below this critical point, a poor compaction of the system is observed. Above $w_c^*$, compaction of the system takes place. The critical point $w_c^*$ indicates a change in the compaction dynamics since around that point the asymptotic density may change of 25\%.  This critical point is shifted towards higher slipping threshold values while the size $L$ of the system is increased.  

The inflexion point corresponding to the critical slipping threshold $w_c^*$ has been estimated by fitting a hyperbolic function in a semi-log scale. Figure 5 presents the critical point $w_c^*$ as a function of the lattice size $L$. The critical slipping threshold $w_c^*$ is found to evolves linearily with the lattice size $L$. According to  
\begin{equation}
\label{tanh2}
w_c^* \approx 0.43 L 
\end{equation} 
as illustrated in Figure 5.
\begin{figure}[htpb] 
\begin{center} 
\includegraphics[width=8cm]{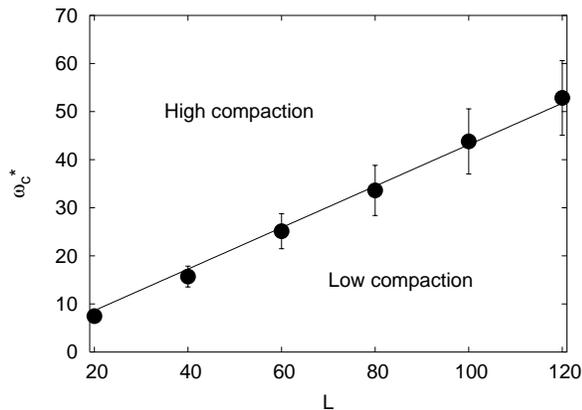}
\caption{The critical slipping threshold $w_c^*$ versus the size $L$ of the lattice. Error bars are indicated. The solid line corresponds to the simple linear fit Eq(\ref{tanh2})}
\end{center} 
\end{figure}
 This linear relationship between $w_c^*$ and $L$ suggests the existence of a characteristic length $\xi$ in the system. This length $\xi$ is thought to be related to the size of the domains of mobile grains. When $\xi$ $<$ $L$, in the packing, only a few grains can move freely since most of the grains are frozen by stable arches. Indeed, when this length grows and reaches the system size $L \approx \xi$, arches touch the borders of the lattice. Then, the numerous grains being part of the arches are frozen in that constrained situation. This occurs at the critical point $w_c^*$. When $\xi$ $>$ $L$, the entire system is fluid.
 
There is thus to a phase transition at the critical point $w_c^*$ which depends on the system size. With the linear ship of Figure 5; the data can be rescaled in order to obtain a unique critical point. This can be seen on Figure 6. The phenomenon is thus independent of the system size $L$. 

\begin{figure}[htpb] 
\begin{center} 
\includegraphics[width=8cm]{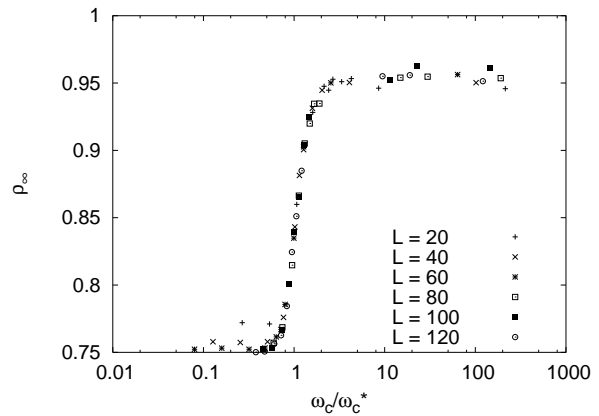}
\caption{Semi-log plot of the asymptotic density $\rho_{\infty}$ as a function of the ratio of slipping threshold $w_c$ and the critical slipping threshold $w_c^*$ for different lattice sizes : $L$ = 20, 40, 60, 80, 100 and 120.}
\end{center} 
\end{figure}

\subsection{Jansen effect}
In this paragraph, the ratio of the total weight supported by the borders is calculated.  The process that allows the weight to be redirected towards the borders is similar to the one uses in the Q-model \cite{hayakawa} : the grains redistribute their weight to their neighbors below them according to some geometrical rules.  In so doing, some stress lines exist and are directed towards the border.  Since a friction is considered with the border, a fraction R of the apparent weight is supported by these later. The fraction $R$ of the apparent weight supported by the border as a function of the slipping threshold $w_c$ is presented in the Figure 7. If the slipping threshold $w_c$ is high enough, there is nearly no friction and the grain cannot transmit their apparent weights to the borders.  

In the Janssen theory of silos \cite{livre}, weights are partially redirected along the horizontal direction. The total pressure $P$ of the packing at the bottom of a silo is given by
\begin{equation}
\label{pressure}
P  = \rho g \chi \left( 1 - \exp\left(-\frac{h}{\chi}\right)\right)
\end{equation} 
where the length $\chi$ includes the effect of friction along the borders. According to this law, we find
\begin{equation}
\chi = \frac{{\chi}_{0} w_c}{w_c*} 
\end{equation} 
where ${\chi}_{0}$ is nearly the grain size. The fraction $R$ of lost weight along the borders is given by 
\begin{equation}
\label{ratio}
R = 1 - \frac{{\chi}_{0} w_c}{hw_c*}\left(1-\exp\left(-\frac{hw_c*}{{\chi}_{0} w_c}\right)\right)
\end{equation} 
which is valid only when the system is not frustated by small arches, i.e $w_c$ $ > $ $w_c^*$. The fraction $R$ tends to zero when the friction is zero ($w_c$$\rightarrow$$\infty$), namely this corresponds to a fluid-like system. From the data of Figure 7, we can conclude that our toy model is in agreement with the Janssen theory for $w_c$ $ > $ $w_c^*$. Below  the critical point $w_c^*$, the physical properties of this frozen system becomes indepent of $w_c$.

\begin{figure}[htpb] 
\begin{center} 
\includegraphics[width=8cm]{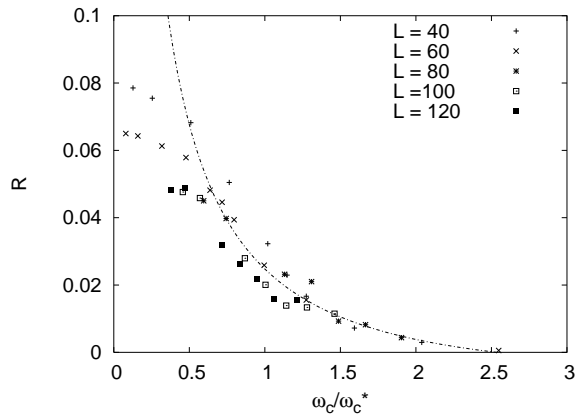}
\caption{Evolution of the ratio $R$ of total pile weight $w$ as a function of the ratio of the slipping threshold $w_c$ and the critical slipping threshold $w_c^*$ for different sizes of the lattice L : 20, 40, 60, 80, 100 and 120. The solid curve is a fit with Eq. \ref{ratio} from $\frac{w_c}{w_c*}$ = 1 to 3.}
\end{center} 
\end{figure}
\subsection{Grain mobility}
According to the previous result that the introduction of the slipping threshold $w_c$ in a simple compaction model has deep effects when regarding asymptotic densities and pressures in silos. How can we measure the influence of friction on the dynamics of compaction ? 

The inverse logarithmic dynamics finds its origin in the geometrical constraints of the model. It has been proposed, as for glassy systems, that the mobility $\mu$ of the grains decreases during compaction, allowing some trapping of defects in the packing. Power laws scaling have been proposed for the grain mobility $\mu$ as a function of $(\rho_{\infty} - \rho)$. Earlier studies \cite{mu1,mu2} have underlined the link between this mobility law and the inverse logarithmic law but the effect of friction has not been studied.
We have numerically measured  $\mu$ in simulations. This quantity $\mu$ is measured during each tap. It is the ratio of the moved grains and the total number of grains in the lowest 25 $\%$ of the lattice. Figure 8 presents a plot of $\mu$ as a funtion of $\rho$ for various values of the slipping threshold $w_c$. The power law scaling 
\begin{equation}
\label{mu}
\mu = \mu_0 + c (\rho_{\infty} - \rho)^\beta
\end{equation}
is found for all data. One should remark that we have considered a residual mobility $\mu_0$ for the free grains in the caverns. The coefficient $c$ being an arbitrary free fitting parameter. The interesting result is that the exponent $\beta$ of the scaling depends clearly on the friction. Below $w_c^*$, the system is frozen and the exponent is close to $\beta \approx 4$. Involving a rapid decrease of the mobilitiy, above $w_c^*$, one has however a low value $\beta \approx \frac{3}{2}$ involving a slow decreases of the mobility (Figure 9). This is remarkable that even when the density does not seem to evolve, the grain mobility is still no-zero.

\begin{figure}[htpb] 
\begin{center} 
\includegraphics[width=8cm]{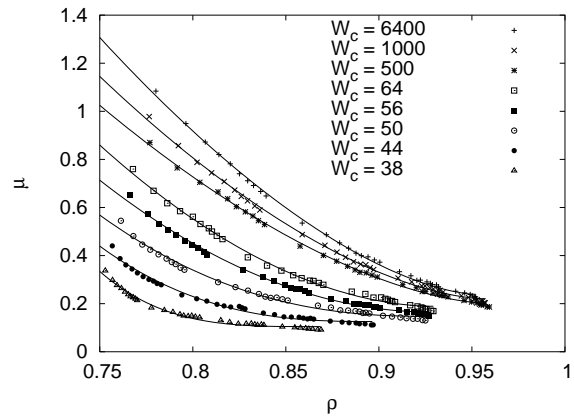}

\caption{Grain mobility $\mu$ as a function of $\rho$ for different value of the slipping threshold $w_c$. The lattice size $L$ is equal to 100.}
\end{center} 
\end{figure}

\begin{figure}[htpb] 
\begin{center} 
\includegraphics[width=8cm]{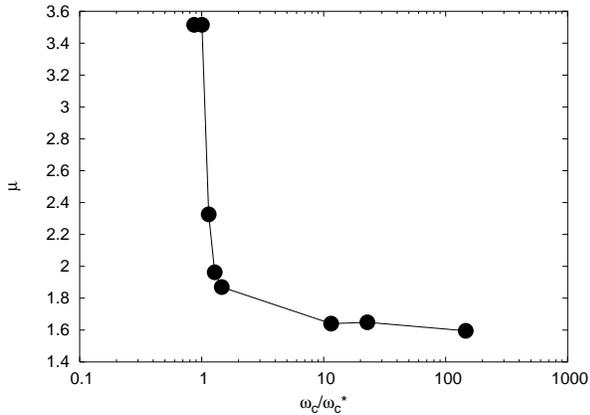}
\caption{Evolution of the exponent $\beta$ as a function of ratio of the slipping threshold $w_c$ and the critical slipping threshold $w_c^*$. The lines conecting dats are only a guide for the eye. The lattice size $L$ is equal to 100.}
\end{center} 
\end{figure}

\section{Conclusion}

Arches have been produced by introducing a slipping threshold $w_c$ related to the static friction coefficient $\mu_s$ between contacting grains.  The density has been studied with respect to the number of taps.  Two main regimes have been found.  For low values of the slipping threshold, the system is completely frozen and the maximum density obtained after tapping is low. For a large value of the threshold $w_c$, the system becomes fluid and a clear compaction is possible like in the Tetris model. The transition between both regime depends on characteristic lengths : some friction length $\xi$ and the system size $L$.

Above the critical point $w_c*$ , the model agrees with the Janssen theory of silos. The ratio $R$ of lost weight on the borders vanishes when $w_c \rightarrow \infty$. 

In order to study the dynamics of compaction, we also measured the mobility of the grains. As proposed in earlier studies \cite{mu1,mu2}, the mobility decreases drastically when $\rho$ reaches the saturation density $\rho_{\infty}$. A power law scaling has been found and the mobility exponent $\beta$ exhibits a large variation above the critical point $w_c^*$.

Our work emphasized the revelance of friction on the dynamics of compaction. Experimental studies are needed in order to test those behaviors.
\section*{Acknowledgements}

 FL benefits of a FRIA grant. SD thanks FNRS for financial support.  This work is financially supported by the ARC contract 02/07-293. We thank H. Caps for fruitful discussions.

\end{document}